# Predicting replicability - analysis of survey and prediction market data from large-scale forecasting projects

Michael Gordon*#1, Domenico Viganola*2, Anna Dreber[3], Magnus Johannesson[3], Thomas Pfeiffer[1]

*The two first authors contributed equally to this work.

#Corresponding author: m.b.gordon@massey.ac.nz

[1]Massey University, Auckland, New Zealand;

[2]George Mason University, Fairfax, VA;

[3]Stockholm School of Economics, Stockholm, Sweden

**Abstract**

The reproducibility of published research has become an important topic in science policy. A number of large-scale replication projects have been conducted to gauge the overall reproducibility in specific academic fields. Here, we present an analysis of data from four studies which sought to forecast the outcomes of replication projects in the social and behavioural sciences, using human experts who participated in prediction markets and answered surveys. Because the number of findings replicated and predicted in each individual study was small, pooling the data offers an opportunity to evaluate hypotheses regarding the performance of prediction markets and surveys at a higher power. In total, peer beliefs were elicited for the replication outcomes of 103 published findings. We find there is information within the scientific community about the replicability of scientific findings, and that both surveys and prediction markets can be used to elicit and aggregate this information. Our results show prediction markets can determine the outcomes of direct replications with 73% accuracy (n=103). Both the prediction market prices and the average survey responses are correlated with outcomes (0.581 and 0.564 respectively, both $p < .001$). We also found a significant relationship between p-values of the original findings and replication outcomes. The dataset is made available through the R package "pooledmaRket" and can be used to further study community beliefs towards replications outcomes as elicited in the surveys and prediction markets.

## 1 Introduction

The communication of research findings in scientific publications plays a crucial role in the practice of science. However, relatively little is known about how reliable and representative the disseminated pieces of information are[1,2]. Concerns have been raised about the credibility of published results following John Ioannidis' landmark essay, "Why most published findings are false"[3], and the identification of a considerable number of studies that turned out to be false positives[4,5]. In response, several large-scale replication projects were initiated in the fields of psychology, experimental economics, and the social sciences more generally[6–14] to systematically evaluate a large sample of findings through direct replication. The rate of successful replication (commonly defined as a result with a significant effect size in the same direction as the original) in these projects ranges from 36% to 62%. This rate, however, cannot be easily compared across projects because key features such as the inclusion criteria and replication power differed across projects. For a discussion of these findings in the context of the 'replication crisis' see refs[1,15–18].

Four of the large-scale replications projects were accompanied by forecasting projects: the Replication Projection: Psychology (RPP)[12]; the Experimental Economics Replication Project (EERP)[6]; the Many Labs 2 Project (ML2)[10]; and the Social Science Replication Project (SSRP)[7]. The replications and the forecasting results were included in a single publication for EERP[6] and SSRP[7]. For RPP [12] and ML2 [10], the forecasting studies appeared separately[19,20]. In each of the replication projects, a set of original findings published in the scientific literature were selected to be repeated via direct replication on new participants and typically larger sample sizes. The purpose of the associated prediction market studies was to investigate whether information elicited from within research communities can be used to predict which findings in the replication projects are likely to replicate; and whether prediction markets and surveys are useful mechanisms for eliciting such information from scientists. The previously published results of the forecasting studies show that the research community can predict which findings are likelihood to replicate – with varying degrees of accuracy. In total, peer beliefs were elicited for the replication outcomes of 103 published findings in the social and behavioural sciences. We have made the resulting dataset available in an R package – 'pooledmaRket'.

In this paper, we present an analysis of this pooled dataset, which allows for both testing hypotheses with substantially higher statistical power and for conducting additional analyses not possible in the previous smaller studies. In the following, we provide a methods section with a brief review of the methodology used in the large-scale replication projects and the prediction market studies as well as how the dataset is analysed in this paper. This is followed by the results of our statistical analysis and a discussion.

# 2 Methods

## 2.1 Replication Projects

Within the replication projects[6,7,10,12], original findings published in the scientific literature were selected based on a set of pre-defined criteria, including research methodology, specific target journals, and time windows. Typically, one key finding of a publication was selected to be replicated with a methodology as close as possible to the original paper. The authors of the original papers were contacted and asked to provide feedback on the replication designs before starting the data collection for the replications.

For RPP, EERP, and SSRP, a replication was deemed successful if it found a 'significant effect size at 5% in the same direction of the original study'[12,21]; for ML2, a replication was deemed successful if it found 'a significant effect size in the same direction of the original study and a p-value smaller than 0.0001'[10]. The latter definition of a successful replication is more stringent because the power of the replications in the ML2 project is higher with multiple laboratories conducting replications. The large-scale replication projects also report additional replication outcomes such as effect sizes.

Statistical power for the replications was typically higher than for the original findings. RPP and EERP had a statistical power of about 90% to find the original effect size. The power was increased substantially for the SSRP project following concerns that effect sizes for original findings may be inflated[12,22], which increases the chances of false negatives among the replication outcomes in the RPP and EERP projects. This was done by using a 2-stage design, where sample sizes in the first stage were such that replications had 90% power to detect 75% of the original effect size. The second stage was conducted if the first stage found a negative result, and together the samples of the two stages led to the replications having 90% power to detect 50% of the original effect size. This two-stage approach is further explained below. In the ML2 study, replications were conducted at multiple sites with large sample sizes, resulting in a substantially higher power.

## 2.2 Forecasting Studies

The four forecasting studies associated with the replication projects investigated the extent to which prediction markets and surveys can be used to forecast the replicability of published findings. Before the replication outcomes became public information, peer researchers participated in a survey eliciting beliefs about the replication probability for findings within the replication projects and thereafter participated in prediction markets.

In the prediction markets, participants were endowed with tokens that could be used to buy and sell contracts each of which paid one token if a finding was replicated, and zero tokens if it was not replicated. At the end of the study, tokens earned from the contracts were converted to monetary rewards. The emerging price for the contracts traded in the market can be interpreted, with some caveats[23], as a collective forecast of the probability of a study replicating. An automated market maker implementing a logarithmic market scoring rule was used to determine prices[24]. The prediction markets were open for two weeks in RPP, ML2, and SSRP, and for 10 days in EERP. The most relevant information for forecasting, including the power of the replications, was embedded in the survey and in the market questions, and the links to the original publications were provided. The forecasters were also provided with the pre-replication versions of the replication reports detailing the design and planned analyses of each replication. In the case of ML2, where many replications were performed for each finding, overall protocols on the replications were provided in lieu of specific replication reports.

Participants were recruited via blogs, mailing lists, and Twitter – with the focus on people working within academia. Some participants who filled out the survey did not participate in the prediction markets. The data presented here is restricted to only those participants who actively participated in the markets, therefore a participant had to trade in at least one market to be included in the survey data.  As a result, both the survey and prediction market data are based on the same participants.

The following subsections provide study specific details; further information is available in the original publications.

### *2.2.1 RPP*

The forecasting study by Dreber et al.[19] was done in conjunction with the Replication Project: Psychology (Open Science Collaboration 2015). In RPP, a sample of findings published in the Journal of Personality and Social Psychology, Psychological Science, and Journal of Experimental Psychology was replicated. The overall replication rate was 36%. The total RPP included 97 original findings, 44 of which were included in both prediction markets and surveys. Dreber et al. ran these 44 prediction markets and 44 surveys in two separate batches in November 2012 and in October 2014 to study whether researchers' beliefs carry useful information about the probability of successful replication. For these 44 studies, 41 replications had been finished at the time of publication. One finding is exluced as it does not have relevant survey forecasts, leaving a total of 40 sets of forecasts included in this dataset. Of the 40 findings included here, prediction markets correctly predicted the outcome of the replications 70% of the time, compared with 58% for the survey. The overall replication rate of the included 40 findings was 37.5% (see Table 1).

#### *2.2.2 EERP*

Camerer et al.[6] replicated 18 findings in the field of experimental economics, published in two of the top economic journals (American Economic Review and Quarterly Journal of Economics). The process for selecting the finding to be replicated from a publication was as follows: (1) select the most central finding in the paper (among the between-subject treatment comparisons) based on to what extent the findings were emphasized in the published versions; (2) if there was more than one equally central finding, the finding (if any) related to efficiency was picked, as efficiency is central to economics; (3) if several findings remained and they were from different separate experiments, the last experiment (in line with RPP) was chosen; (4) if several findings still remained, one of those findings was randomly selected for the replication. A fraction of 61% of replications were successful. Both the markets and the survey correctly categorized 11 findings out of 18 (61%).

#### *2.2.3 ML2*

Forsell et al.[25] presents forecasts for replications included in the ML2 project[10], a further large-scale replication project led by the Open Science Collaboration. One of the aims of the ML2 project was to guarantee high-quality standards for the replications of classic and contemporary findings in psychology by using large sample sizes across different cultures and laboratories and requiring replication protocols to be peer-reviewed in advance. The findings were selected by the authors of the ML2 project, with the aim of assuring diversity and plurality of findings. The realized replication rate for the ML2 project was 46% (11 successful replications out of 24 findings analysed). Although ML2 replicated in total 28 findings, replication outcomes were only forecasted for 24 of these. The excluded findings focused on cultural differences in effect sizes across different samples. Note that when including all 28 findings, the replication rate of the Many Labs 2 project[10] increases to 50% (14/28). Further detail is given in Appendix A in Forsell et al.[20]. The prediction markets correctly predicted 75% of the replication outcomes. As a comparison, the survey correctly predicted 67% of replication outcomes.

#### *2.2.4 SSRP*

SSRP is a replication project covering 21 experimental social science studies published in two high-impact interdisciplinary journals, Science and Nature[7]. SSRP was specifically designed to address the issue of inflated effect sizes in original findings. There were three criteria for selecting findings within publications (presented in descending order): (1) select the first finding that reports a significant effect; (2) select the statistically significant finding identified as the most important; (3) randomly select a single finding in cases of more than one equally central result. In line with previous studies, Camerer et al [7], also ran prediction markets and prediction surveys to forecast whether the selected studies will replicate. The design of SSRP for conducting replications differed from the previous projects in that it was structured in two stages: the first stage provided 90% power to detect 75% of the original effect size; if the replication failed, stage 2 started and the data collection continued until there was 90% power of detecting 50% of the original effect size (pooling data from the stage 1 and stage 2 collection phases). Based on all the data collected, 62% of the 21 findings were successfully replicated. The prediction markets followed a similar structure of the data collection: participants were randomized in two groups: in treatment 1, beliefs about replicability in stage 1 were elicited; in treatment 2, beliefs about replicability in both stage 1 and stage 2 were elicited. In this paper, we report the results about treatment 2 only, as the replication results after Stage 2 are the most informative about the replication outcome.

### 2.3 Pooled Dataset

Due to the high similarity in research topic and design of the four forecasting studies, they can be pooled into a single dataset. The pooled data can be downloaded within the R package which can be accessed at https://github.com/MichaelbGordon/pooledmaRket. The dataset is presented in three separate tables, combined from the four forecasting studies as well as a codebook which provides details on each of the columns within each dataset. Each table represents the key parts of the studies; replication outcomes and original findings features, survey responses, and prediction market trades. Each of these tables is made available in the R package, as well as example code of aggregation methods.

In order to analyse the performance of the prediction markets we typically take the market price at time of closing as the aggregated prediction of the market. For the survey we aggregate primarily with simple mean, but also provide performance of several other aggregations. In total we analyse data from over 15,000 forecasts across the 103 findings, made up of 7850 trades and 7380 survey responses.

## 3 **Results**

In this section, we report and comment on the outcomes of the descriptive and statistical analyses performed to compare the prediction markets results and the survey results. For all the results reported below, the tests are interpreted as two-tailed tests and a p-value < 0.005 is interpreted as "statistically significant" while a p-value < 0.05 as "suggestive" evidence, in line with the recommendation of Benjamin et al. [26].

### 3.1 Observed and Forecasted Replication Rates

Successful rates of replication ranged from 38% to 62%, with an overall rate of 49% for the 103 findings included in the dataset (Table 1). The Replication Project Psychology had the lowest overall replication rate of 38%. Many Labs 2 had the second-lowest replication rate with 11 out of 24 studies successfully replicating (45.8%). The Experimental Economics Replication Project and Social Studies Replication Project both have replication rates around 60%.

Expected replication rates as found within prediction markets range from 56% (for RPP) to 75% (for EERP). Surveys predicted replication rates from 55% (RPP) and 71% (EERP). Overall, the replication rates as expected by the community is around 60%. When comparing actual with expected replication rates, both the average survey ($M$ = 0.61, $SD$ = 0.14) and final market price ($M$=0.63, $SD$=0.21) tend to overestimate the actual rate of replication success ($M$ = 0.49). Paired t tests found statistically significant difference between actual replication rate and the survey ($t(102) = -2.89$), $p = 0.0046$) and the market ($t(102) = -3.43$, $p = 0.00088$).

The binary outcome variable is correlated with both the survey responses ($r(101) = .564$, $p < .001$) and market prices($r(101) = .581$, $p < .001$), as shown in Figure. Market based and survey based forecasts in all four studies are highly correlated (RPP - $r_s(38) = 0.736$, $p < .001$; EERP - $r_s(16)$ =0 .792, $p < .001$; SSRP - $r_s(19) = 0.845$, $p < .001$; ML2 - $r_s(22) = 0.947$, $p < .001$). When considering combined data the same high correlation is found ($r_s(101) = .837$, $p < .001$; $r(101) = .853$, $p < .001$); as it emerges distinctly from Figure 2.

## 3.2 Accuracy of forecasts

In order to assess the effectiveness of forecasters providing beliefs via prediction markets and surveys, we analyse error rates for each method, and overall accuracy when adopting a binary approach. For the binary approach we interpret a final price of 0.50 or greater as prediction of a successful replication and a final price lower than 0.50 as prediction of a failed replication. The same rules applied for surveys: we computed the mean beliefs for each study and then interpreted that the survey predicts a successful replication if the average beliefs exceed 0.50 and a failed replication otherwise. Using this approach, the surveys never outperformed the markets. In two cases (EERP and SSRP) they correctly categorize the same number of findings in the replicates/non-replicates dichotomy; in the other two studies, the markets do better (71% vs 58% in the RPP; 75% vs 67% in the ML2). Overall, the prediction markets had an accuracy of 73% (75 out of 103 studies), while the surveys had an accuracy of 66% (68 out of 103 studies). However, based on a chi-square test this difference is not statistically significant ($X^2$ (1) = 1.12, p = 0.29).

Findings that do not replicate tend to have prediction market prices below the 0.50 threshold, while studies that do replicate are more concentrated above the 0.5 threshold. Out of the 31 findings that are predicted by the market not to replicate, only three eventually replicated, thus for these findings the market is correct more than 90% of the time. Alternatively, 25 of the 73 (66%) findings that were predicted to replicate did not. The survey-based predictions follow a similar pattern; of the 22 findings that are predicted to not replicate by the survey, only two eventually replicated and of the 81 studies that are predicted to successfully replicate, 33 did not replicate. Both the market and survey-based forecasts are more accurate when concluding that a study will not replicate rather than when concluding that a study will replicate, markets ($X^2$ (1) = 6.68, p = 0.01; $X^2$ (1) = 4.45, p = 0.035 for markets and surveys respectively). This may at least be partially due to the limited power of the replications in RPP and EERP, as some of the failed replications may be false negatives.

The absolute error, defined as the absolute difference between actual replication outcome (either 0 or 1) and the forecasted chance of replication, is used as an accuracy measure that does not entail a loss of information from binarizing the aggregated forecasts. The forecasts for SSRP had the lowest mean absolute error of 0.303 and 0.348 for the prediction markets and survey respectively. This was followed by ML2 (market error of 0.354 and survey error of 0.394) and RPP (market error of 0.431 and survey error of 0.485). Only in EERP there was a lower absolute mean error for the survey – 0.409 compared with 0.414 for the market. Across all 103 findings, the absolute error of the prediction markets (*M*

= 0.384) is significantly lower than the survey (*M* = 0.423) ($t(102) = 3.68$, p = 0.0003). Prediction markets tend to provide more extreme forecasts with the final price ranges being larger in all four projects than the survey beliefs range. Quantifying extremeness as distance of a forecast from 0.5 the markets show a significantly larger extremeness compared to the average survey ($t(102) = 7.87$, p <0.0001). Overall, market-based forecasts and survey-based forecasts are similar when using a binary metric, however the more extreme market forecasts provide a significantly better predictor when evaluating based on error.

## 3.3 Aggregation Methods

Using alternate aggregations of the individual survey response can create more extreme forecasts which have been linked to better forecasts[27]. We provide results for two additional survey aggregation methods; median, simple voting and variance weighted mean. Simple voting includes binarizing every survey response (effectively rounding each response to either 0 or 1) and reporting the percentage of responses which vote for replication success. Variance weighted mean is based on finding a positive relationship between variance in survey responses and overall accuracy. We hypothesize that forecasters with a large variance in survey responses are able to better discriminate between which studies are likely to replicate and which aren't likely, thus providing more extreme forecasts. On the other hand, forecasters who are not able to discriminate provide similar forecasts for many studies, and therefore have a low variance. The median aggregator (*M* = 0.63, *SD* = 0.17), the simple voting aggregator (*M*= 0.66, *SD* = 0.21) and the variance weighted mean (*M* =58, *SD* = 17) methods provided higher variance forecasts than the mean (*M* = 0.610, *SD* = 0.14). Evaluating the survey aggregations using mean absolute error simple voting performs the best (0.39), followed variance weighted mean (0.407) and median (0.412). The mean aggregator has the highest mean absolute error of 0.422. The final market price still outperforms these alternate aggregations.

## 3.4 Market Dynamics

Predictions market are designed to aggregate information that is widely dispersed amongst agents. The market price is expected to converge to a relatively stable value which is interpreted as a probability of the outcome occurring [28,29]. For replication markets it is unknown how quickly the market can converge. Using both time and number of trades to quantify how the market progresses, we can investigate on average at what point the information distributed across the agents is 'priced into the market'.

The number of trades in each of the markets range from 26 to 193 ($M = 76$, $SD = 32$) we observe that each forecasting study opened their prediction markets for between 11 and 14 days. The market trades tend to be front loaded, where trading activity diminishes over the available trading time.

As expected the markets experience diminishing returns in terms of reduction in mean absolute error. Reductions in mean absolute error can be modelled using LOESS regression, where the mean absolute error is estimated at different numbers of trades[30]. This model shows that 90% error reduction (i.e 90% of the total error reduction that will occur) happens in the first 69 trades.

When analysing error reduction as a function of time, 65% of the error reduction that will be achieved occurs in the first hour. 90% of total error reduction occurs within the first 161 hours of the markets (just under a week). Both in terms of number of trades and time, the average error fluctuates towards the end the market, without consistently improving forecasts, indicating trades made towards the end of the markets are noisy (figure 3). However, applying a time weighting smoothing algorithm of all trades after the first week does not result in a significant increase in accuracy.

3.5 P value Analysis

The p-values of the findings has been shown to be correlated with the replication outcomes [20,21]. In particular two other replication based forecasting attempts has shown that p-values are informative to a machine learning algorithm[31,32]. We here test this relationship using the pooled market data. One limitation of this analysis is that p-values are often reported as an inequality or a category rather than a real number, for example a typical reported p value is "$p < 0.05$". Therefore, we transform p-values into categories. As a guide we use categories as suggested by Benjamin et al. [26]; of 'suggestive evidence, $p > 0.005$, and statistical significance $p <= 0.005$. This two-category approach provides a significant relationship between strength of evidence (through p-values) and replication outcomes ($b = 0.46$, $p < .001$). While the replication rate for findings with $p <= 0.005$ is about 74%, the replication rate for findings with $p > 0.005$ drops to 28%. The correlation of p-value category and outcomes is 0.456 (given by R-squared of the linear model). Full results can be found in Table 2.

## 4 **Discussion**

In this paper, we investigate the forecasting performances obtained by two different procedures to elicit beliefs about replication of scientific studies: prediction markets and prediction survey. We pooled the forecasting data using these two methods from four published papers in which forecasters, mainly researchers and scholars in the social sciences, had to estimate the probability that a tested hypothesis taken from a paper published in scientific journals would replicate. We find that the prediction markets correctly identify replication outcomes 73% of the times (75/103), while the prediction surveys are correct 66% of the times (68/103). Both the prediction market estimates and the prediction surveys estimates are highly correlated with the replication outcomes of the studies selected for replication (Pearson correlation = 0.581 and = 0.564, respectively), suggesting that studies that replicate can be distinguished from studies that do not successfully replicate. However, both the forecasts elicitation methods tend to overestimate the realized replication rates, and beliefs about replication are on average about ten percentage units larger than the observed replication rate. The results suggest that peer beliefs can be elicited to obtain important information about reproducibility, but the systematic overestimation of the replication probability also imply that there is room for calibrating the elicited beliefs to further improve predictions.

In terms of comparing which elicitation method performs better in the task of aggregating beliefs and providing more accurate forecasts, our results suggest that the markets perform somewhat better than the survey especially if evaluating based on absolute prediction error. We confirmed previous results which indicated that p-values are informative in respect to replication success. This also provides some context for interpreting p-values at different levels as strength of evidence. The data and results found this paper can be used for the future forecasting projects either planned or in progress[14], by informing experimental design and forecasting aggregation. The pooled dataset presents opportunities for other researchers investigate replicability of scientific research, human forecasts and their intersection, as well as providing a benchmark for any further replication-based markets.

**Authors' contributions**

All authors contributed to drafting the manuscript. M.G collated and formatted the data, and built the R package. M.G. and D.V. carried out the statistical analyses.

**Competing interests**

We declare we have no competing interests.

**Acknowledgements**

T.P. thanks the Marsden Fund for financial support for project MAU-1710. We are also grateful for a grant from the Jan Wallander and Tom Hedelius Foundation (Svenska Handelsbankens Forskningsstiftelser) to Anna Dreber. Anna Dreber is funded by the Knut and Alice Wallenberg Foundation and the Marianne and Marcus Wallenberg Foundation (through a Wallenberg Scholar grant) and the Austrian Science Fund (FWF, SFB F63). Anna Dreber, Magnus Johannesson, and Domenico Viganola were also funded by the Jan Wallander and Tom Hedelius Foundation (Svenska Handelsbankens Forskningsstiftelser) and the Swedish Foundation for Humanities and Social Sciences. This material is based upon work supported by the Defense Advanced Research Projects Agency (DARPA) and Space and Naval Warfare Systems Center Pacific (SSC Pacific) under contract no. N66001-19-C-4014.

## 5 Bibliography

1. Baker, M. 1,500 scientists lift the lid on reproducibility. *Nat. News* **533**, 452 (2016).
2. John, L. K., Loewenstein, G. & Prelec, D. Measuring the Prevalence of Questionable Research Practices With Incentives for Truth Telling. *Psychol. Sci.* **23**, 524–532 (2012).
3. Ioannidis, J. Why Most Published Research Findings Are False. *PLOS Med.* **2**, e124 (2005).
4. Ioannidis, J. & Doucouliagos, C. What's to Know About the Credibility of Empirical Economics? *J. Econ. Surv.* **27**, 997–1004 (2013).
5. Maniadis, Z., Tufano, F. & List, J. A. One Swallow Doesn't Make a Summer: New Evidence on Anchoring Effects. *Am. Econ. Rev.* **104**, 277–290 (2014).
6. Camerer, C. F. *et al.* Evaluating replicability of laboratory experiments in economics. *Science* **351**, 1433–1436 (2016).
7. Camerer, C. F. *et al.* Evaluating the replicability of social science experiments in Nature and Science between 2010 and 2015. *Nat. Hum. Behav.* **2**, 637–644 (2018).

8. Ebersole, C. R. *et al.* Many Labs 3: Evaluating participant pool quality across the academic semester via replication. *J. Exp. Soc. Psychol.* **67**, 68–82 (2016).
9. Klein, R. A. *et al.* Investigating variation in replicability: A "many labs" replication project. *Soc. Psychol.* **45**, 142–152 (2014).
10. Klein, R. A. *et al.* Many Labs 2: Investigating Variation in Replicability Across Samples and Settings. *Adv. Methods Pract. Psychol. Sci.* **1**, 443–490 (2018).
11. Landy, J. *et al.* Crowdsourcing hypothesis tests: Making transparent how design choices shape research results. *Psychol. Bull.* (2019).
12. Open Science Collaboration. Estimating the reproducibility of psychological science. *Science* **349**, aac4716 (2015).
13. Schweinsberg, M. *et al.* The pipeline project: Pre-publication independent replications of a single laboratory's research pipeline. *J. Exp. Soc. Psychol.* **66**, 55–67 (2016).
14. Gordon, M. *et al.* Are replication rates the same across academic fields? Community forecasts from the DARPA SCORE programme. *R. Soc. Open Sci.* **7**, 200566.
15. Christensen, G. & Miguel, E. Transparency, Reproducibility, and the Credibility of Economics Research. *J. Econ. Lit.* **56**, 920–980 (2018).
16. Etz, A. & Vandekerckhove, J. A Bayesian Perspective on the Reproducibility Project: Psychology. *PloS One* **11**, e0149794 (2016).
17. Fanelli, D. Opinion: Is science really facing a reproducibility crisis, and do we need it to? *Proc. Natl. Acad. Sci.* **115**, 2628–2631 (2018).
18. Pashler, H. & Harris, C. R. Is the Replicability Crisis Overblown? Three Arguments Examined. *Perspect. Psychol. Sci.* **7**, 531–536 (2012).
19. Dreber, A. *et al.* Using prediction markets to estimate the reproducibility of scientific research. *Proc. Natl. Acad. Sci.* **112**, 15343–15347 (2015).

20. Forsell, E. *et al.* Predicting replication outcomes in the Many Labs 2 study. *J. Econ. Psychol.* **75**, 102117 (2019).
21. Cumming, G. Replication and p Intervals: p Values Predict the Future Only Vaguely, but Confidence Intervals Do Much Better. *Perspect. Psychol. Sci.* **3**, 286–300 (2008).
22. Ioannidis, J. P. A. Why Most Discovered True Associations Are Inflated: *Epidemiology* **19**, 640–648 (2008).
23. Manski, C. F. Interpreting the predictions of prediction markets. *Econ. Lett.* **91**, 425–429 (2006).
24. Hanson, R. Combinatorial Information Market Design. *Inf. Syst. Front.* **5**, 107–119 (2003).
25. Forsell, E. *et al.* Predicting replication outcomes in the Many Labs 2 study. *J. Econ. Psychol.* (2018) doi:10.1016/j.joep.2018.10.009.
26. Benjamin, D. J. *et al.* Redefine statistical significance. *Nat. Hum. Behav.* **2**, 6 (2018).
27. Baron, J., Mellers, B. A., Tetlock, P. E., Stone, E. & Ungar, L. H. Two Reasons to Make Aggregated Probability Forecasts More Extreme. *Decis. Anal.* (2014) doi:10.1287/deca.2014.0293.
28. Arrow, K. J. *et al.* The Promise of Prediction Markets. *Science* **320**, 877–878 (2008).
29. Atanasov, P. *et al.* Distilling the Wisdom of Crowds: Prediction Markets vs. Prediction Polls. *Manag. Sci.* **63**, 691–706 (2017).
30. Cleveland, W. S. & Devlin, S. J. Locally Weighted Regression: An Approach to Regression Analysis by Local Fitting. *J. Am. Stat. Assoc.* **83**, 596–610 (1988).
31. Yang, Y., Youyou, W. & Uzzi, B. Estimating the deep replicability of scientific findings using human and artificial intelligence. *Proc. Natl. Acad. Sci.* **117**, 10762–10768 (2020).
32. Altmejd, A. *et al.* Predicting the replicability of social science lab experiments. *PLOS ONE* **14**, e0225826 (2019).

## 6 Figures & Tables

**Table 1: Main features of individual projects.** This table contains key characteristics and summaries of the datasets and the pooled data. In calculations of correct forecasts by the prediction market and survey, we interpreted a final price of 0.50 or greater as prediction of a successful replication; if the final price is lower than 0.50, we interpret this as prediction of a failed replication. Overall the actual replication rate was 49%, indicating that the forecasters were over confident with the average market price being 0.627. Prediction Markets tend to outperform survey's when forecasting replication success when considering overall accuracy – 73% compared to 66%.

| | **RPP** | **EERP** | **ML2** | **SSRP** | **Pooled data** |
|---|---|---|---|---|---|
| **Replication Study** | Ref [12] | Ref [6] | Ref [10] | Ref [7] | |
| **Forecasting Study** | Ref [19] | | Ref [25] | | |
| **Field of study** | Experimental Psychology | Experimental Economics | Experimental Psychology | Experimental Social Science | |
| **Source Journals** | JPSP, PS, JEP (2008) | AER, QJE (2011-2014) | Several psychology outlets, including JEP, JPSP, PS (1977-2014) | Science, Nature (2010-2015) | |
| **Replicated Findings** | 40 | 18 | 24 | 21 | 103 |
| **Successful replications** | 15 (37.5%) | 11 (61.1%) | 11 (45.8%) | 13 (61.9%) | 51 (49%) |
| **Mean beliefs - Prediction Market** | 0.556 | 0.751 | 0.644 | 0.634 | 0.627 |
| **Correct – Prediction Markets (%)** | 28(70%) | 11 (61%) | 18 (75%) | 18 (86%) | 76 (73%) |
| **Mean Absolute Error – Prediction Market** | 0.431 | 0.414 | 0.354 | 0.303 | 0.384 |
| **Mean beliefs - survey** | 0.546 | 0.711 | 0.647 | 0.605 | 0.610 |
| **Correct - Survey (%)** | 23 (58%) | 11 (61%) | 16 (67%) | 18 (86%) | 68 (66%) |
| **Spearman Correlation – Prediction Market and Survey beliefs** | 0.736 | 0.792 | 0.947 | 0.845 | 0.837 |
| **Spearman Correlation – Replication Outcomes and Prediction Market** | 0.418 | 0.297 | 0.755 | 0.842 | 0.568 |
| **Spearman Correlation – Replication Outcomes and Survey beliefs** | 0.243 | 0.516 | 0.731 | 0.760 | 0.557 |

**Table 2. P-Value and Replication success.** This table includes the details of a linear model regressing p-value category and replication outcomes. We show that studies with p-values < 0.005 are replicated at 45 percentage points more often than the reference category - ‘p > 0.005’.

Association between P value (2 categories) and observed replication outcomes.

| | *Dependent variable:* |
|---|---|
| | *Replication Outcome (binary [0,1])* |
| Intercept | 0.2807** |
| | (0.0595) |
| P value < 0.005 | 0.458** |
| | (0.0890) |
| Observations | 103 |
| $R^2$ | 0.2079 |

*Note:* $*p < 0.05$; $**p < 0.005$. Standard errors in brackets.

**Figure 1. Market Beliefs.** This figure plots the final prices of the 103 markets included within this dataset ordered by price. The green dots represent successful replications and the non-replications are represented by the red dots. The horizontal line at 0.5 indicates the binary cut off used to determine the markets aggregated prediction.

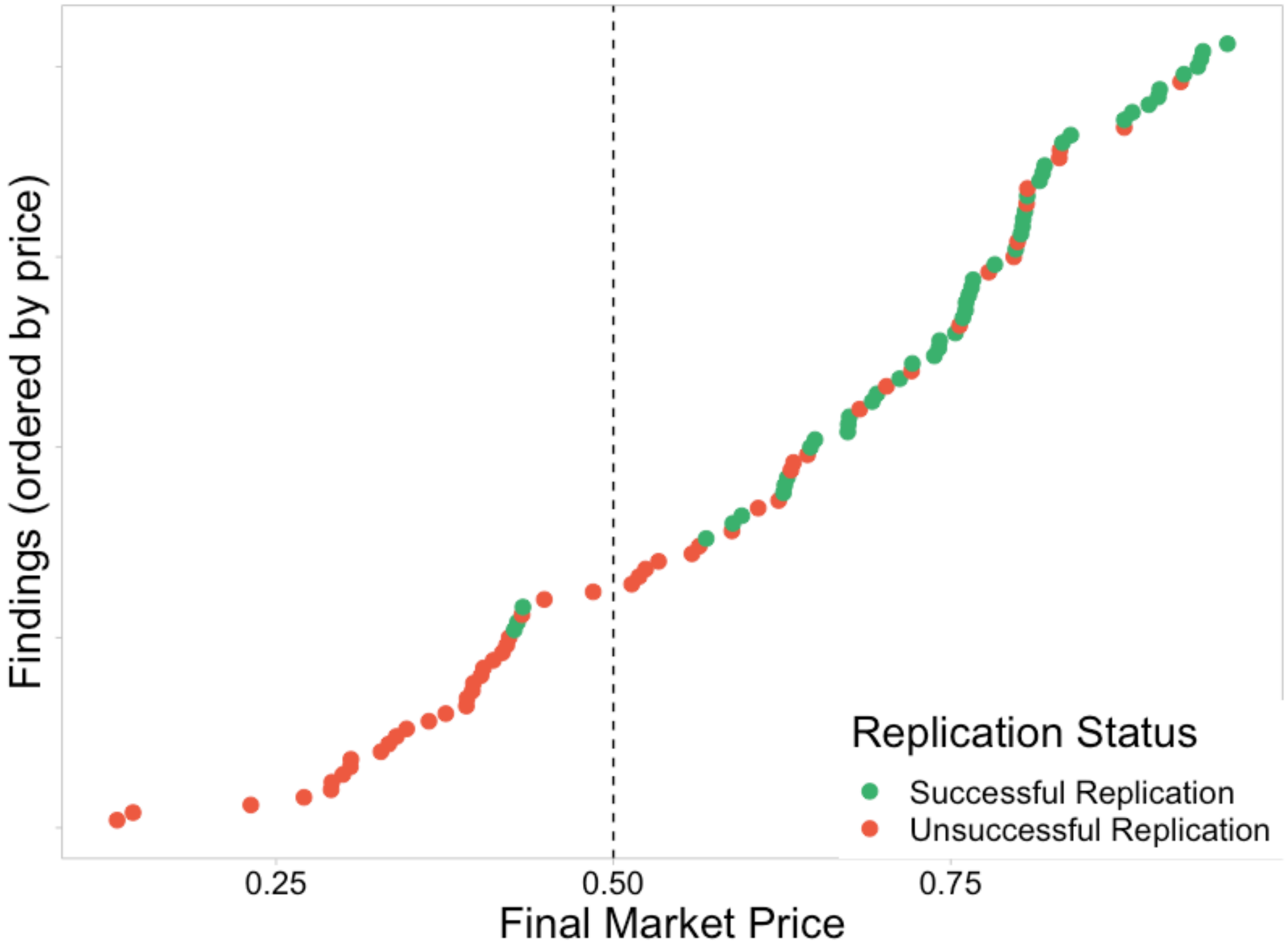

**Figure 2. Market and Survey Correlations.** Final Market Prices and average survey responses are highly correlated (rs(101) = .837, p < .001; r(101) = .853, p < .001). The dotted horizontal and vertical lines indicate the 0.5 cut off points used when applying a binary forecasting approach. The top left quadrant represents those findings which are predicted to replicate by survey but predicted to not replicate by market. The top right and bottom left quadrants contain findings where the markets and surveys agree, predicted to replicated and to not replicate respectively. The bottom right quadrant with a single finding, is where the study is predicted to replicate by the market but not by the survey. The colours of the findings show the replication outcome, with green indicating a successful replication outcome, and red indicating unsuccessful replication.

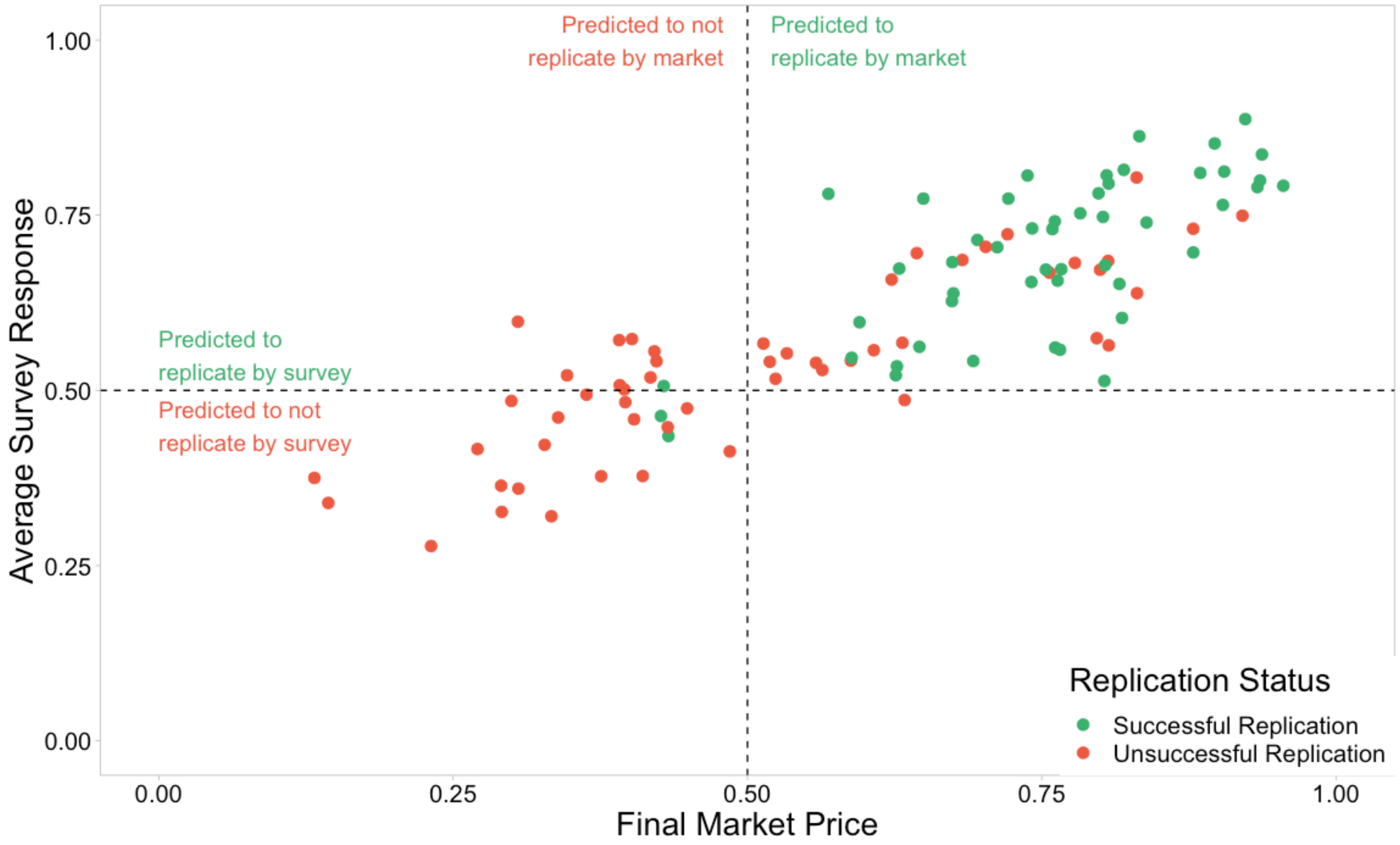

**Figure 3. Market Dynamics.** Each of the subplots represent reduction in absolute error as the market progresses. Both plots include the mean (across 103 findings) absolute error in black, and the LOESS smoothing in blue. Plot (a) describes error reduction over number of trades and plot (b) describes error reduction over time. We find that the error falls quickly at first, however error does not reduce over

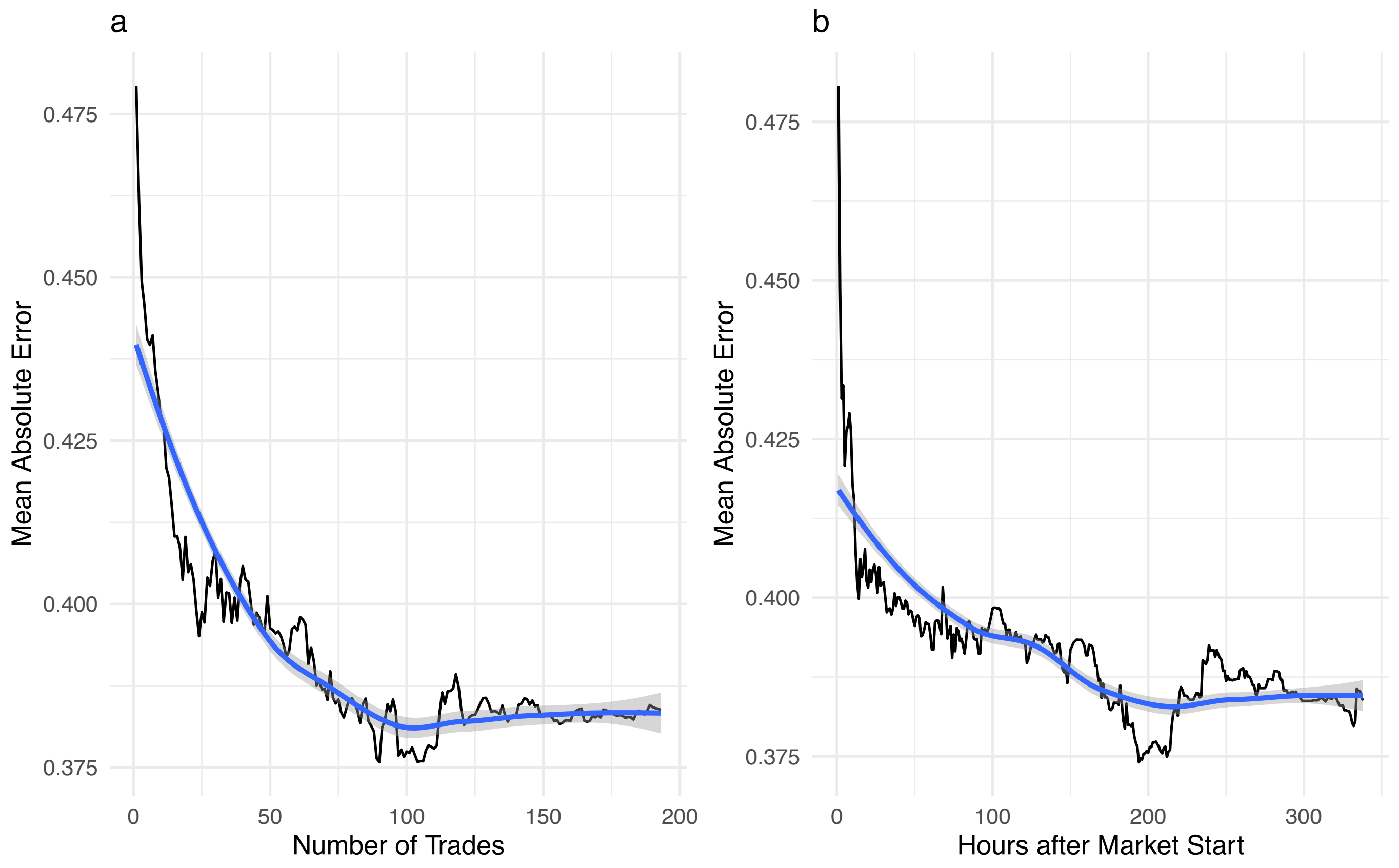